\documentclass{cimento}
\usepackage{graphicx,color}

\title{Threshold-cusp explanation for 
$X$ and $Z_{cs}$ in $B^+\to J/\psi\phi K^+$
}
\author{S.X.~Nakamura\from{ins:x}\thanks{satoshi@sdu.edu.cn}\ETC 
 \atque
X.~Luo\from{ins:y}
\instlist{\inst{ins:x} Institute of Frontier and Interdisciplinary Science, Shandong
University - Qingdao, China 
  \inst{ins:y} 
School of Physics and Optoelectronics Engineering, Anhui University -
Hefei, China
}}

\begin{document}

\maketitle

\begin{abstract}
Several $X$ and $Z_{cs}$ exotic hadrons were claimed in the
LHCb's amplitude analysis on $B^+\to J/\psi \phi K^+$.
The data shows that 
all the peaks and also dips in the spectra
are located at thresholds of seemingly
relevant meson-meson channels.
While the LHCb analysis fitted the peaks with
Breit-Wigner resonances,
threshold kinematical cusps might be the cause of such structures.
We thus analyze the LHCb data considering the threshold cusps.
Our model is simultaneously fitted to
$J/\psi \phi$, $J/\psi K^+$, and 
$K^+\phi$ invariant mass distributions.
The threshold cusps fit well
all the $X$, $Z_{cs}$, and dip structures. 
Our analysis indicates that 
spin-parity of the 
$X(4274)$ and $X(4500)$ structures are $0^-$ and $1^-$, respectively.
This is different from the LHCb's spin-parity assignments ($1^+$ and $0^+$).
The number of fitting parameters can be significantly reduced by
considering the relevant threshold cusps.
Our analysis shows that 
$D_s^{(*)}\bar{D}^{*}$ scattering lengths are consistent with zero.
This disfavors an explanation of $Z_{cs}(4000)$ and $Z_{cs}(4220)$ as
$D_s^{(*)}\bar{D}^{*}$ molecules, which is 
consistent with lattice QCD via the SU(3) relation.
\end{abstract}

\section{Introduction}

The LHCb's six-dimensional amplitude analysis for $B^+\to J/\psi \phi K^+$
found several $X$ and $Z_{cs}$ exotic states
that are different from the conventional 
$q\bar{q}$~\cite{lhcb_phi1,lhcb_phi2,lhcb_phi}.
Those appearing as bumps in the $J/\psi\phi$ invariant mass
($M_{J/\psi\phi}$) distribution are
$X(4140)$ and $X(4274)$ with $J^P=1^+$, and 
$X(4500)$ and $X(4700)$ with $J^P=0^+$ ($J$: spin, $P$: parity).
Also, $Z_{cs}(4000)^+$ and $Z_{cs}(4220)^+$ with $J^P=1^+$
appear as bumps in the $M_{J/\psi K^+}$ distribution.
The mass, width, and $J^P$ are primarily important for studying
the nature and structures of the exotic hadrons.
However, due to assumptions and simplifications,
amplitude analysis results are sometimes not unique nor model-independent.

A basic assumption in the LHCb's amplitude analysis is that all the bumps 
are caused by resonances that can be well simulated by the Breit-Wigner form.
However, the $X$ bumps are located at $D_s^*\bar{D}_s^{(*)}$,
$D_{sJ}^{(*)}\bar{D}_s^{(*)}$, and $\psi'\phi$ thresholds, 
and the $Z_{cs}$ bumps at $D_{s}^{(*)}\bar{D}^*$ thresholds.
There are also dips at other thresholds.
Since resonancelike and dip structures might have been generated by 
threshold cusps, 
the LHCb's assumption of using the Breit-Wigner amplitudes should be
viewed with caution. 
The assumption would also influence the $J^P$ assignments.
The LHCb's $J^P$ assignments might change if 
threshold cusps at the peaks and dips are taken into account in fitting
the data.
It seems important to conduct an independent amplitude analysis for the
same process with due consideration of the threshold cusps,
and compare with the LHCb's results.

The nature of the $Z_{cs}(4000)$ and $Z_{cs}(4220)$ has been a
controversial issue.
The BESIII collaboration observed a similar structure, $Z_{cs}(3985)$,
in $e^+ e^- \to K^+(D_s^- D^{*0}+D_s^{*-} D^0)$~\cite{BESIII:2020qkh}.
Their masses are similar: 
$4003\pm 6^{+4}_{-14}$~MeV for $Z_{cs}(4000)$ and
$3982.5^{+1.8}_{-2.6}\pm 2.1$~MeV for $Z_{cs}(3985)$.
However, their widths are rather different:
$131\pm 15\pm 26$~MeV for $Z_{cs}(4000)$ and
$12.8^{+5.3}_{-4.4}\pm 3.0$~MeV for $Z_{cs}(3985)$.
Several proposals have been made to understand whether
$Z_{cs}(3985)$ and 
$Z_{cs}(4000)$ are
the same or different,
and whether they are
$cu\bar{c}\bar{s}$ tetraquark state
or $D_s^{(*)}\bar{D}^{(*)}$ molecules.
It is also possible that 
a virtual pole enhances 
the $D_s\bar{D}^{*}$ threshold cusp
to generate the $Z_{cs}(3985)$ and $Z_{cs}(4000)$ structures.
$Z_{cs}(3985/4000)$ and $Z_{c}(3900)$~\cite{bes3_zx3900}
may be SU(3) partners.
The existence of a narrow $Z_{c}(3900)$ state is disfavored
by Lattice QCD (LQCD) calculations, implying that
$Z_{c}(3900)$ may be a kinematical effect.
Via the SU(3) relation, the LQCD is not in favor of the existence of 
a narrow $Z_{cs}(3985/4000)$ state.

In this work, we develop a reaction model for $B^+ \to J/\psi \phi K^+$~\footnote{
See Ref.~\cite{full_paper} for the full account of this work.
}.
Then, the model is fitted to the LHCb data of 
the ${J/\psi\phi}$, ${J/\psi K^+}$, and ${K^+\phi}$ invariant mass
distributions simultaneously.
We show that 
ordinary $s$-wave threshold cusps from one-loop diagrams in Fig.~\ref{fig:diag}
can fit all the peaks and dips in the 
$M_{J/\psi\phi}$ and $M_{J/\psi K^+}$ distributions,
and no virtual poles near the thresholds are necessary.
Our model indicates that 
the $X(4274)$ and $X(4500)$ structures are from the 
$D_{s0}^*(2317)^+{D}^-_s$ and 
$D_{s1}(2536)^+{D}^-_s$ threshold cusps that have 
$J^P=0^-$ and $1^-$, respectively.
This is different from the LHCb's $J^P$ assignments 
($J^P=1^+$ and $0^+$), which would originate from 
quite different mechanisms in the models.
Possible advantages of our model over the LHCb's model will be pointed
out. 
Furthermore, we examine whether 
the LHCb data favors 
the $D_s^{(*)}\bar{D}^{*}$ molecule interpretation of 
$Z_{cs}(4000)$ and $Z_{cs}(4220)$.
Our analysis will show that 
the data requires 
the $D_s^{(*)}\bar{D}^{*}$ scattering lengths to be consistent with
zero within our model.
Thus the molecule interpretation is disfavored.
This is actually consistent with the above-mentioned 
LQCD implication.

\section{Model}
\begin{figure}[t]
\begin{center}
\includegraphics[width=0.25\textwidth]{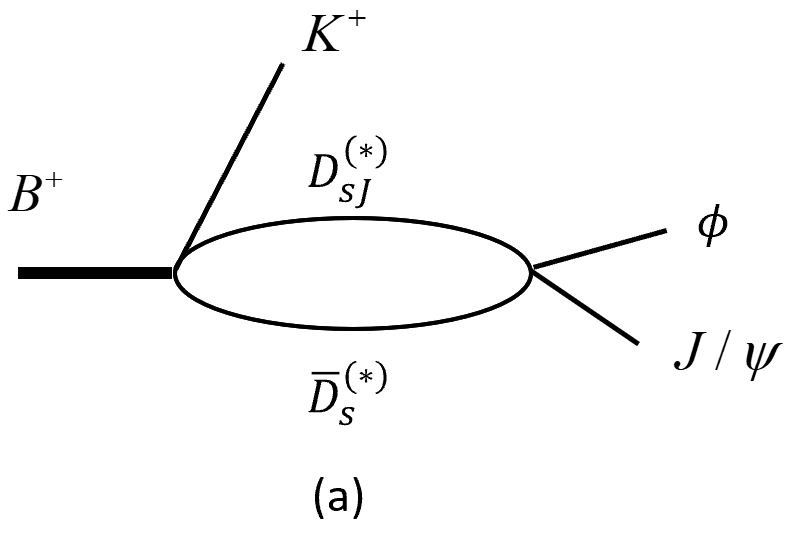}
\includegraphics[width=0.25\textwidth]{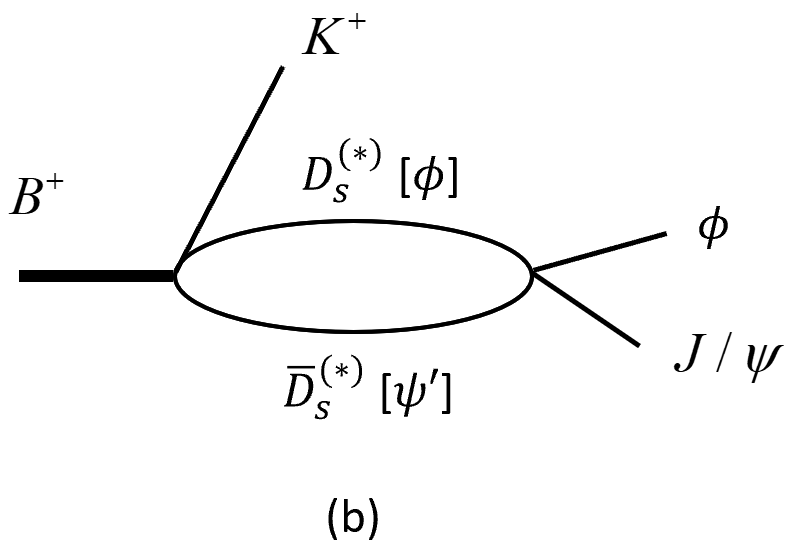}
\includegraphics[width=0.25\textwidth]{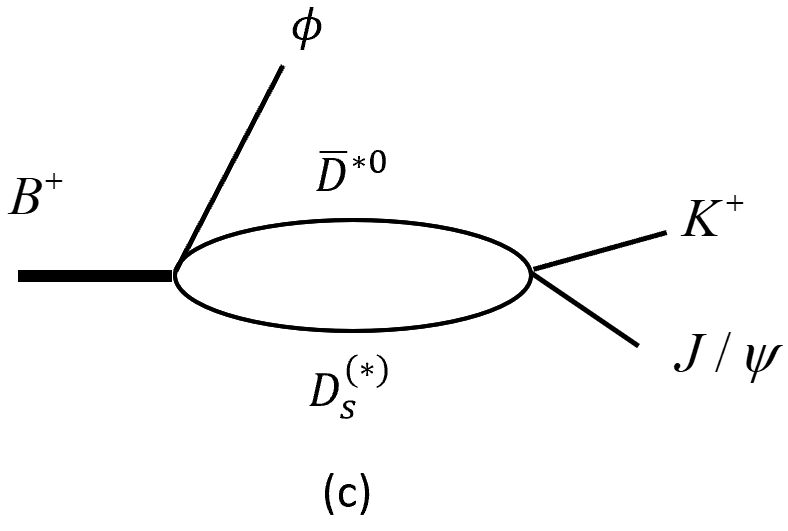}
\includegraphics[width=0.20\textwidth]{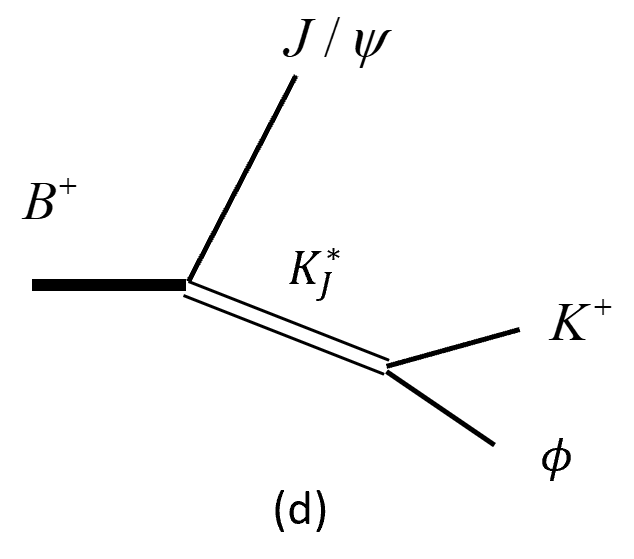}
\end{center}
 \caption{Diagrams for
$B^+\to J/\psi \phi K^+$ considered in this work.
(a) one-loop of $D_{sJ}^{(*)+}{D}_s^{(*)-}$ ($0^-, 1^-$);
(b) one-loop of $D_{s}^{*+}{D}_s^{(*)-}$ and $\psi'\phi$ ($0^+, 1^+$);
(c) one-loop of $D_{s}^{(*)+}\bar{D}^{*0}$ ($1^+$);
(d) $K^*_J$ ($K$, $K^*, K_1$, $K_2$) resonances.
$D_{sJ}^{(*)}$ refers to $D_{s0}^*(2317)$ and $D_{s1}(2536)$,
and $\psi'$ is $\psi(2S)$.
Figures taken from Ref.~\cite{full_paper}. Copyright (2021) APS.
 }
\label{fig:diag}
\end{figure}

For describing the
structures in the $M_{J/\psi\phi}$ and $M_{J/\psi K^+}$
distributions of $B^+\to J/\psi \phi K^+$ with threshold cusps,
one-loop mechanisms of 
Figs.~\ref{fig:diag}(a-c) are considered.
$K^*_J$ resonance mechanisms of Fig.~\ref{fig:diag}(d) are also
considered and they would 
shape the $M_{K^+\phi}$ distribution.
In the one-loop mechanisms of Fig.~\ref{fig:diag}(a),
$s$-wave pairs of
$D_{s0}^*(2317)^+{D}^-_s(0^-)$,
$D_{s0}^*(2317)^+{D}^{*-}_s(1^-)$,
$D_{s1}(2536)^+{D}^-_s(1^-)$, and
$D_{s1}(2536)^+{D}^{*-}_s(0^-)$
are included;
$J^P$ is indicated in the parenthesis.
In one-loop mechanisms of Fig.~\ref{fig:diag}(b),
$s$-wave pairs of
$D_{s}^{*+}{D}^-_s(1^+)$,
$D_{s}^{*+}{D}^{*-}_s(0^+)$,
$\psi'\phi(0^+)$, and
$\psi'\phi(1^+)$
are included.
In Fig.~\ref{fig:diag}(c), 
$s$-wave pairs of
$D_s^+\bar{D}^{*0}(1^+)$ and
$D_s^{*+}\bar{D}^{*0}(1^+)$ 
are included in the one-loop.
Several $K^*_J$ resonance mechanisms in Breit-Wigner forms
are considered in Fig.~\ref{fig:diag}(d).

An amplitude formula for Fig.~\ref{fig:diag}(c) 
with a $D_{s}^+ \bar{D}^{*0}(1^+)$ pair
is given below as a representative case. 
The energy, width, three-momentum and 
polarization vector of a particle $x$ 
are denoted by
$E_x$, $\Gamma_x$, $\vec p_x$ and $\vec \varepsilon_x$, respectively.
We refer to the PDG~\cite{pdg} for
the particle masses and widths.
The amplitudes includes 
a $B^+ \to D_s^+ \bar{D}^{*0}\phi$ vertex, which is parity-conserving (pc), 
and a $D_s^+\bar{D}^{*0}\to J/\psi K^+$ interaction.
Their matrix elements are
\begin{eqnarray}
&c_{D_s\bar{D}^{*0} (1^+)}^{\rm pc}\vec \varepsilon_{\bar{D}^{*0}} \cdot \vec \varepsilon_\phi F^{00}_{D_s\bar{D}^{*0} \phi,B},
\qquad
&c_{\psi K, D_s\bar{D}^{*0}}^{1^+} \vec \varepsilon_{\bar{D}^{*0}}
 \cdot \vec \varepsilon_{\psi} f_{\psi K}^0 f^0_{D_s\bar{D}^{*0}},
\end{eqnarray}
respectively,
where $c_{D_s\bar{D}^{*0} (1^+)}^{\rm pc}$ and 
$c_{\psi K, D_s\bar{D}^{*0}}^{1^+}$
are coupling constants.
We have introduced dipole form factors denoted by
$F_{ijk,l}^{LL'}$ and $f_{ij}^{L}$ where 
a common cutoff of $\Lambda=1$~GeV is used in all form factors.
Combining the above matrix elements,
the amplitude for Fig.~\ref{fig:diag}(c) with $D_{s}^+ \bar{D}^{*0}(1^+)$
is given as
\begin{eqnarray}
\label{eq:zcs4000amp}
A_{\bar{D}^{*0}D_s(1^+)}^{\rm 1L,pc} &=&
c_{\psi K, D_s\bar{D}^{*0}}^{1^+}
c_{D_s\bar{D}^{*0} (1^+)}^{\rm pc}
\vec \varepsilon_\psi \cdot \vec \varepsilon_\phi 
 \int d^3 p_{D_s} {
f_{\psi K}^0 f^0_{D_s\bar{D}^{*0}}F^{00}_{D_s\bar{D}^{*0}
\phi,B}\over 
M_{\psi K}-E_{D_s}-E_{\bar{D}^{*0}}+i\varepsilon}.
\end{eqnarray}
The product of coupling constants 
($c_{\psi K, D_s\bar{D}^{*0}}^{1^+}c_{D_s\bar{D}^{*0} (1^+)}^{\rm pc}$),
which is a complex overall factor, 
is determined by fitting the LHCb data.
In our default fit,
we use 16 mechanisms in total, giving 
$2\times 16 -3 = 29$ fitting parameters.
Since 
the absolute normalization of the full amplitude and 
overall phases of the parity-conserving and -violating full amplitudes
are arbitrary, we subtracted 3 from the number of the fitting
parameters.

There might be 
virtual or bound states near the 
$D_s^{(*)+} \bar{D}^{*0}$ thresholds, and 
they could enhance 
the threshold cusps
from Eq.~(\ref{eq:zcs4000amp}).
This effect can be implemented in our model by 
describing the $D_s^{(*)+} \bar{D}^{*0}\to J/\psi K^+$ transition
with a single-channel $D_s^{(*)+} \bar{D}^{*0}$ scattering, 
and a perturbative 
$D_s^{(*)+} \bar{D}^{*0}\to J/\psi K^+$ transition follows.
Our $D_s^{(*)+} \bar{D}^{*0}$ interaction potential is
\begin{eqnarray}
v_\alpha (p',p) &=& f^0_\alpha(p') h_\alpha\; f^0_\alpha(p) ,
\label{eq:cont-ptl}
\end{eqnarray}
where an interaction channel is labeled by $\alpha$ and 
the coupling constant is $h_\alpha$.
The rescattering effect on the $Z_{cs}$ structures are not included in
our default fit, but it will be studied separately.

\begin{figure}[t]
\begin{center}
\includegraphics[width=0.495\textwidth]{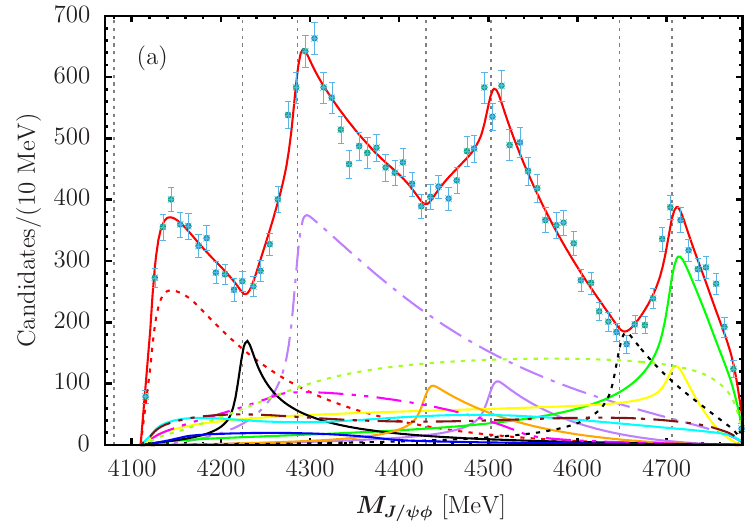}
\includegraphics[width=0.495\textwidth]{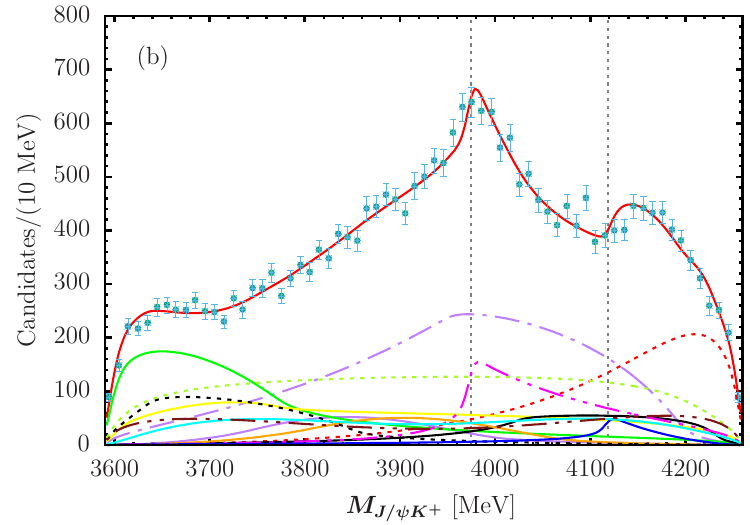}
\end{center}
\caption{\label{fig:res}
(a) $J/\psi \phi$ and (b) $J/\psi K^+$
invariant mass distributions for $B^+ \to J/\psi \phi K^+$.
The data are from LHCb~\cite{lhcb_phi}.
Our default fit is shown by the red solid curves.
Each loop contribution of Fig.~\ref{fig:diag}(a-c) is shown such as
$D_{s}^{*+}{D}^-_s(1^+)$ by red-dashed,
$D_{s}^{*+}{D}^{*-}_s(0^+)$ by black-solid,
$D_{s0}^*(2317)^+{D}^-_s(0^-)$ by purple-dash-dotted,
$D_{s0}^*(2317)^+{D}^{*-}_s(1^-)$ by orange-solid,
$D_{s1}(2536)^+{D}^-_s(1^-)$ by purple-solid,
$D_{s1}(2536)^+{D}^{*-}_s(0^-)$ by black-dashed,
$\psi'\phi(0^+)$ by green-solid,
$\psi'\phi(1^+)$ by yellow-solid,
$D_s^+\bar{D}^{*0}(1^+)$ by magenta-dash-two-dotted, and
$D_s^{*+}\bar{D}^{*0}(1^+)$ by blue-solid.
Also, each contribution from Fig.~\ref{fig:diag}(d) is shown:
$K^*$ by cyan-solid, $K_1$ by green-dashed, and $K_2$ by brown-dash-two-dotted.
The dotted vertical lines indicate threshold locations.
From left to right, 
$D_{s}^*\bar{D}_s$,
$D_{s}^*\bar{D}^*_s$,
$D_{s0}^*(2317)\bar{D}_s$,
$D_{s0}^*(2317)\bar{D}^*_s$,
$D_{s1}(2536)\bar{D}_s$,
$D_{s1}(2536)\bar{D}^*_s$, and
$\psi'\phi$ thresholds for the panel (a), and
$D_s^+\bar{D}^{*0}$ and $D_s^{*+}\bar{D}^{*0}$
thresholds for the panel (b).
Figures taken from Ref.~\cite{full_paper}. Copyright (2021) APS.
}
\end{figure}

\section{Results and discussions} 

The LHCb data for $B^+\to J/\psi \phi K^+$ such as 
the $M_{J/\psi\phi}$, $M_{J/\psi K^+}$, 
and $M_{K^+\phi}$ distributions
are simultaneously fitted with 
the above-described model, 
and the fit result is shown in 
Fig.~\ref{fig:res}.
The LHCb data are well fitted by our default model.
The fit quality can be quantified by
$\chi^2/{\rm ndf}=1.77$ 
where ndf is the number of degrees of freedom (the number of 
bins subtracted by the number of fitting parameters).

The peaks in the $M_{J/\psi\phi}$ distribution
are well fitted by the threshold cusps such as:
the $X(4140)$ peak by the $D_{s}^{*+}{D}^-_s(1^+)$ threshold cusp [red dashed curve],
$X(4274)$ by $D_{s0}^*(2317)^+{D}^-_s(0^-)$ [purple dash-dotted], 
$X(4500)$ by $D_{s1}(2536)^+{D}^-_s(1^-)$ [purple solid], and 
$X(4700/4685)$ by $\psi'\phi(0^+/1^+)$ [green solid/yellow solid].
Also, the three dips are generated by 
the corresponding threshold cusps located at
$D_{s}^{*+}{D}^{*-}_s(0^+)$ [black solid],
$D_{s0}^*(2317)^+{D}^{*-}_s(1^-)$ [orange solid], and
$D_{s1}(2536)^+{D}^{*-}_s(0^-)$ [black dashed]
thresholds.

For the $M_{J/\psi K^+}$ distribution, 
the $Z_{cs}(4000)$-like peak is well described by 
the $D_s^+\bar{D}^{*0}(1^+)$ threshold
cusp shown by the magenta dash-two-dotted curve.
The $Z_{cs}(4220)$-like structure is also formed by our model. 
However, it is not from a resonance but from 
 the $D_s^{*+}\bar{D}^{*0}(1^+)$ threshold
cusp [blue solid] appearing as a dip and 
the shrinking phase-space near the kinematical endpoint.

There are several qualitative differences between 
our and LHCb's analysis results.
Our model describes all $X$ and $Z_{cs}$ structures 
with the threshold cusps, while the LHCb's model uses 
the Breit-Wigner-type resonances.
Our model also has threshold cusps for the dips in the 
$M_{J/\psi\phi}$ spectrum.
On the other hand, the LHCb's model uses interferences to generate the dips.
Probably due to the differences in the mechanisms, 
the $J^P$ assignments to the $X(4274)$ and $X(4500)$ peaks are also different:
$0^-$ for $X(4274)$ and $1^-$ for $X(4500)$ in our model, but
$1^+$ for $X(4274)$ and $0^+$ for $X(4500)$ in the LHCb's.

The number of fitting parameters ($N_{\rm p}$) is also rather different
between our and LHCb's models.
Our default model has $N_{\rm p}=29$ to fit
the $M_{J/\psi\phi}$, $M_{J/\psi K^+}$, 
and $M_{K^+\phi}$ distributions, while 
the LHCb's amplitude model 
has $N_{\rm p}=144$ to fit the six-dimensional distribution.
Since the LHCb's model is fitted to the richer information, 
$N_{\rm p}$ should be larger to some extent.
Nevertheless, this alone does not seems to fully explain the larger $N_{\rm p}$ 
in the LHCb's model.
A possible explanation for the large $N_{\rm p}$ is that
relevant mechanisms are missed in the LHCb's model, 
and this needs to be compensated for by including many other mechanisms and
complicated interferences.

In the default fit,
the $Z_{cs}$-like structures 
in the $M_{J/\psi K^+}$ distribution
are well described by
the threshold cusps and no nearby poles are necessary.
Does the LHCb data allow 
a molecule (pole) scenario for the $Z_{cs}$ structures ?
We address this question by varying the fitting parameters involved in
Fig.~\ref{fig:diag}(c), and also 
$D_s^{+} \bar{D}^{*0}$ and $D_s^{*+} \bar{D}^{*0}$
 interaction strengths $h_\alpha$ in Eq.~(\ref{eq:cont-ptl});
$h_\alpha$ for the $D_s^{+} \bar{D}^{*0}$ interaction and
that for $D_s^{*+} \bar{D}^{*0}$ can take different values.
Then we obtain the allowed regions for the two $h_\alpha$ values.
We obtain the regions for 
the $D_s^{+} \bar{D}^{*0}$ interaction strength as 
$-0.33 < h_\alpha < 0.93$, corresponding to 
the scattering length of 
$-0.12 < a {\rm (fm)} < 0.06$.
This indicates that a virtual pole exists at
93~MeV below the $D_s^{+} \bar{D}^{*0}$ threshold or deeper.
Similarly, 
we obtain the regions for 
the $D_s^{*+} \bar{D}^{*0}$ interaction strength as 
$-0.17 < h_\alpha < 2.02$, corresponding to 
$-0.21 < a {\rm (fm)} < 0.03$, and 
a virtual pole exists at 103~MeV below the threshold or deeper.
This result indicates that the LHCb data does not favorably support 
the interpretation that 
the $Z_{cs}$ structures are $D_s^{(*)+} \bar{D}^{*0}$ molecules.

The previous LQCD calculations for the $J^{PC}=1^{+-}$ $D^*\bar{D}$ scattering
indicated that the interaction is weak, and that the scattering would
not form
a bound or narrow resonance corresponding to $Z_c(3900)$.
Via the SU(3) relation, the LQCD results implies that 
the $Z_{cs}(4000)$ structure in the LHCb data would not be due to a $D_s^{+} \bar{D}^{*0}$
molecule.
Our default model as well as the analysis result in the previous
paragraph are consistent with this LQCD finding. 
Phenomenologically, however, 
the $Z_c(3900)$~\cite{bes3_zx3900} and
$Z_{cs}(3985)$~\cite{BESIII:2020qkh} peak structures in the data have
not been well explained by a non-pole scenario.
Experimental, phenomenological, and LQCD studies are more needed 
to draw a consistent picture for $Z_{c(s)}$.

\acknowledgments
SXN is supported by 
National Natural Science Foundation of China (NSFC) under contracts 
U2032103.
XL is supported by the National Natural Science Foundation of China
under Grant No. 12205002.

\end{document}